\documentclass[conference,a4paper]{IEEEtran}
\usepackage{enumerate}
\usepackage[top=0.8in, bottom=1.5in, left=0.6in, right=0.6in]{geometry}
\usepackage{cite}
\usepackage{amscd}
\usepackage{dsfont}
\usepackage{latexsym}
\usepackage{array}

\usepackage{tikz}\usetikzlibrary{calc}
\usepackage{tabularx}
\usepackage{epsfig}
\usepackage{graphicx,subfigure}
\usepackage{amsmath,mathrsfs}
\usepackage{amsthm}
\usepackage{amssymb}  %used for the font style \mathbb{}
\usepackage{amsbsy}
\usepackage{color}
\usepackage{stfloats}
\usepackage{algorithm}
\usepackage{algpseudocode}
\usepackage{bm}
\usepackage{pifont}
\usepackage{accents}
\usepackage{multirow}

\newtheorem{lem}{Lemma}
\newtheorem{ther}{Theorem}
\newtheorem{deft}{Definition}

\theoremstyle{definition}
\newtheorem{rem}{Remark}

\ifCLASSINFOpdf
\else
\fi
\begin{document}
\title{Construction of Simultaneously Good Polar Codes and Polar Lattices}

\author{
\IEEEauthorblockN{Ling Liu$^1$, Ruimin Yuan$^2$, Shanxiang Lyu$^3$, Cong Ling$^4$, Baoming Bai$^1$}
\IEEEauthorblockA{$^1$Guangzhou Institute of Technology, Xidian University, Guangzhou, China}
\IEEEauthorblockA{$^2$State Key Lab. of ISN, Xidian University, Xi’an, China}
\IEEEauthorblockA{$^3$College of Cyber Security, Jinan University, Guangzhou, China}
\IEEEauthorblockA{$^4$Department of Electrical and Electronic Engineering, Imperial College London, London, UK}
\IEEEauthorblockA{liuling@xidian.edu.cn, ruiminyuan@stu.xidian.edu.cn, lsx07@jnu.edu.cn, cling@ieee.org, bmbai@mail.xidian.edu.cn}
}


%Telephone: (800) 555--1212, Fax: (888) 555--1212}


\maketitle

\begin{abstract}
In this work, we investigate the simultaneous goodness of polar codes and polar lattices. The simultaneous goodness of a lattice or a code means that it is optimal for both channel coding and source coding simultaneously. The existence of such kind of lattices was proven by using random lattice ensembles. Our work provides an explicit construction based on the polarization technique.
\end{abstract}


% \fi
\IEEEpeerreviewmaketitle

\section{Introduction}
The simultaneous goodness of lattices originates from the celebrated work of \cite{zamir2}, where Erez et al. proved that there exists a sequence of lattices $\Lambda_n$ of increasing dimension $n$, which is good for covering, quantization, packing and modulation simultaneously. More explicitly, as $n \to \infty$, the covering radius satisfies $\rho_{\rm{cov}}(\Lambda_n) \to 1$, the normalized second moment (NSM) satisfies $G(\Lambda_n) \to 1/2\pi e$, the packing radius satisfies $\lim \inf \rho_{\rm{pack}}(\Lambda_n) \geq \frac{1}{2}$, and meanwhile the normalized volume to noise ratio (NVNR) satisfies $\mu(\Lambda_n, P_e) \to 2\pi e$ for any error probability $P_e >0$. From the perspective of information theory, the first two properties of lattices are closely related to lossy source coding, while the latter two are highly connected with channel coding problem, especially for the additive white Gaussian noise (AWGN) channel. The duality between the first two properties (for quantization) and the latter two (for transmission) was studied in \cite{yellowbook,ForneyDual93}. More details on the lattice properties can be found in \cite{BK:Zamir}.

 Except for duality, there also exists certain ordering relationship among these properties. In fact, the covering-goodness entails quantization-goodness, and the packing-goodness is stronger than modulation-goodness in the sense that the former corresponds to hard packing while the latter corresponds to soft packing. This work is less ambitious than \cite{zamir2} as we will mainly focus on quantization-goodness and modulation-goodness, which are typically adequate to solve many information-theoretical problems. In contrast to the random ensemble of lattices in \cite{zamir2}, our task is to provide an explicit construction method for lattices that are simultaneously quantization good and modulation good. Such lattices can be used in lattice based quantization index modulation \cite{ChenWornellQIM,SXQIM21} and some scenarios where source coding and channel coding are jointly designed \cite{VerduJSCC10}.

Our simultaneously good lattices are constructed from binary polar codes using the Construction D method \cite{yellowbook}. Since their invention, polar codes have been consecutively shown to be good for channel coding \cite{arikan2009channel} and source coding \cite{KoradaSource}. Notice that the constructions of polar codes for these two problems are slightly different, and a good polar code for one problem is not necessarily good for the other one. To overcome this issue, we propose a chaining framework based on two polar codes with identical coding rates. The resulting polar code, with a longer block length, can be proved to be good both for channel and source coding. We note that the chaining technique is similar to that used for universal polar codes in compound channels \cite{PolarUniHassani}. However, the difference lies in the fact that the construction of universal polar codes requires several ingredient polar codes separately designed for different channels but with the same capacity, whereas the simultaneously good polar codes are based on two ingredient polar codes designed for two different channels with distinct capacities. As a result, more careful treatment is needed and the parallel universal polarization technique \cite{PolarUniSasoglu} may not be applied.

Based on the simultaneously good polar codes, we then construct simultaneously good polar lattices, following the multilevel structure given in \cite{forney6}. This approach has successfully generated quantization-good polar lattices \cite{QZgoodITW2024} and modulation-good polar lattices \cite{polarlatticeJ}. The new constructed lattice is expected to build a bridge between the two existing good lattices, and it also provides a geometric instance of the discrete Gaussian shaping, which has been shown to be optimal for both modulation \cite{LingBel13} and quantization \cite{CongQZ}.

The remainder of the paper is organized as follows. \mbox{Sec. II} gives preliminaries of polar codes and lattices. The simultaneously good polar codes are described in \mbox{Sec. III}. In \mbox{Sec. IV}, we demonstrate how such codes can be used to build good lattices. The two goodnesses will be briefly proved, and the paper is concluded in \mbox{Sec. V}.

$\it{Notation:}$ All random variables (RVs) are denoted by capital letters. Let $P_X$ denote the probability mass function of a RV $X$ taking values in a countable set $\mathcal{X}$, and the probability density function (PDF) of $Y$ in an uncountable set $\mathcal{Y}$ is denoted by $f_Y$. The combination of $N$ i.i.d. copies of $X$ is denoted by a vector $X^{1:N}$ or $X^{[N]}$, where $[N]=\{1,...,N\}$, and its $i$-th element is given by $X^i$. When $N$ is clear from the context, we write $X^{[N]}$ as $\mathbf{X}$ for brevity. The realization of $X^{[N]}$ ($\mathbf{X}$) is given as $x^{[N]}$ ($\mathbf{x}$). The subvector of $\mathbf{X}$ with indices limited to a subset $\mathcal{F} \subseteq [N]$ is denoted by $X^{\mathcal{F}}$. The cardinality of $\mathcal{F}$ is $|\mathcal{F}|$, and its complement set is $\bar{\mathcal{F}}$. A lattice partition chain is indicated by $\Lambda (\Lambda_0)/\Lambda_1/\cdots/\Lambda_i/\cdots$, where $\Lambda_i$ is a one-dimensional (1-D) lattice. When chaining $B$ coding blocks, a set of indices in the $j$-th block is denoted by $\mathcal{I}^j$ for $1\leq j\leq B$. We use $\log$ for the binary logarithm and $\ln$ for the natural logarithm. $\lceil\cdot\rceil$ ($\lfloor\cdot\rfloor$) denotes the ceiling (flooring) function. Standard asymptotic notation is used throughout this paper.

\section{Preliminaries of Polar Lattices}\label{sec:background}

\subsection{Polar Codes for Channel and Source Coding}
For the case of channel coding, denote by $W$ a binary-input memoryless symmetric channel (BMSC) with uniform input $X \in \mathcal{X}$ and output $Y \in \mathcal{Y}$. Its channel transition probability is given by $P_{Y|X}$. The Shannon capacity of $W$ is denoted by $I(W)$. Let $N=2^m$ for some positive integer $m$. The polarization matrix is defined as
\begin{eqnarray}
\mathbf{G}_N \triangleq \left[\begin{matrix}1&0\\1&1\end{matrix}\right]^{\otimes m} \times \mathbf{B}_N,
\end{eqnarray}where $\otimes$ denotes the Kronecker product, and $\mathbf{B}_N$ is the bit-reverse permutation matrix \cite{arikan2009channel}. The matrix $\mathbf{G}_N$ transforms the inputs $X^{[N]}$ of $N$ identical copies of $W$ to $U^{[N]}=X^{[N]}\mathbf{G}_N$, which gives a vector channel $W_N: U^{[N]}\to Y^{[N]}$. $W_N$ can be successively split into $N$ binary memoryless symmetric synthetic channels $U^i\to (U^{1:i-1}, Y^{[N]})$, denoted by $W_{N}^{(i)}$ with $1 \leq i \leq N$. The core of channel polarization states that, as $N$ increases, $W_N^{(i)}$ polarizes to a good (roughly error-free) channel or a totally noisy one almost surely. Moreover, the fractions of these two extreme synthetic channels turn to $I(W)$ and $1-I(W)$, respectively. To achieve the capacity, one can choose a rate $R_c < I(W)$, transmit information bits through $\lfloor R_cN \rfloor$ good synthetic channels, and feed frozen bits (pre-shared to the receiver before transmission) to the rest ones.

For the case of source coding, $X$ and $Y$ are generally termed the reconstruction RV and the source RV, and $W$ is called the test channel between them. The performance of lossy compression is measured by a distortion function $\mathrm{d}: \mathcal{Y}\times\mathcal{X}\to \mathbb{R}_+$, which is naturally extended to the vector form as $\mathrm{d}(y^{[N]},x^{[N]}) =\sum_{i=1}^N \mathrm{d}(y^i,x^i)$. In this work, we choose the Hamming distortion function $\mathrm{d}(y,x) = y \oplus x$ for i.i.d. Bernoulli sources and the Euclidean distortion function $\mathrm{d}(y,x) = (y-x)^2$ for i.i.d. Gaussian sources. Shannon's rate-distortion theorem \cite{ShannonRD1959} characterized the optimal compression rate $R(D)$ for a target average distortion $D$ as the lowest mutual information between $X$ and $Y$, i.e.,
\begin{eqnarray}
R(D) = \min_{P_{X|Y}: \mathbb{E}_P[\mathrm{d}(y,x)]\leq D} I(X;Y).
\end{eqnarray} Note that for a given source distribution $P_{Y}$, the above optimal $P_{X|Y}$ defines the $P_{Y|X}$ of the test channel $W$. Korada and Urbanke \cite{KoradaSource} showed that for symmetric $Y$ and $W$, $R(D)=I(W)$ can be achieved by polar codes as follows. First we perform channel polarization according to $\mathbf{G}_N$ in the same way as the channel coding case. The reconstruction vector is converted to $U^{[N]}= X^{[N]}\mathbf{G}_N$. For a bad synthetic channel, its input $U^i$ is almost independent of its output $(Y^{[N]},U^{1:i-1})$. For a given rate $R_s > I(W)$, the compression of $Y^{[N]}$ is done by freezing the inputs of the bad synthetic channels and then determining the rest $\lceil R_sN \rceil$ bits according to $Y^{[N]}$ using the successive cancellation (SC) decoding method.

In practice, the indices of good or bad synthetic channels can be identified on the basis of their associated Bhattacharyya parameters.
\begin{deft}\label{deft:symZ&asymZ}
Given a BMSC $W$ with transition probability $P_{Y|X}$, the Bhattacharyya parameter of $W$ is defined as
\begin{eqnarray}
Z(W)=Z(X|Y)\triangleq\sum\limits_{y} \sqrt{P_{Y|X}(y|0)P_{Y|X}(y|1)}.
\end{eqnarray}
\end{deft}

In \cite{arikan2009rate}, the rate of polarization $\beta$ is analyzed, which characterizes how fast $Z(W_N^{(i)})$ approaches 0 or 1. For the 2-by-2 kernel $\left[\begin{smallmatrix}1&0\\1&1\end{smallmatrix}\right]$, $\beta$ is upper bounded by $\frac{1}{2}$. That is, for any $0<\beta<\frac{1}{2}$, we have
\begin{eqnarray}
\lim_{N\to\infty} P \left(Z(W_{N}^{(i)}) <2^{-N^{\beta}}\right) = I(W),
\end{eqnarray}and
\begin{eqnarray}
\lim_{N\to\infty} P \left(Z(W_{N}^{(i)}) >1-2^{-N^{\beta}}\right) = 1-I(W).
\end{eqnarray}
As a result, the information set can be chosen from $\{i\in [N]:Z(W_{N}^{(i)})< 2^{-N^{\beta}}\}$ for channel coding, while the frozen set can be chosen from $\{i\in [N]:Z(W_{N}^{(i)}) > 1-2^{-N^{\beta}}\}$ for source coding.

\begin{deft}\label{deft:Degrade}
Let $W_1:X \to Y_1$ and $W_2: X \to Y_2$ be two channels. $W_1$ is stochastically degraded with respect to (w.r.t.) $W_2$ if there exists an intermediate channel $W:Y_2\rightarrow Y_1$ such that
\begin{eqnarray}
W_1(y_1|x)=\sum_{y_2\in\mathcal{Y}_2}W_2(y_2|x)W(y_1|y_2).
\end{eqnarray}
\end{deft}

Given two BMSCs $W$ and $V$, if $V$ is degraded w.r.t. $W$, then $Z(W_N^{(i)}) \leq Z(V_N^{(i)})$ for $i\in [N]$ after channel polarization \cite{polarchannelandsource}.

\subsection{Lattice Codes and Polar Lattices}
An $n$-dimensional lattice is a discrete subgroup of $\mathbb{R}^{n}$ which can be described by
\begin{eqnarray}
\Lambda=\{ \bm{\lambda}=\mathbf{z}\mathbf{B}:\mathbf{z}\in\mathbb{Z}^{n}\},
\end{eqnarray}
where the rows of the generator matrix $\mathbf{B}=[\mathbf{b}_{1}; \cdots; \mathbf{b}_{n}]$ are assumed to be linearly independent.

For a vector $\mathbf{x}\in\mathbb{R}^{n}$, the nearest-neighbor quantizer associated with $\Lambda$ is $Q_{\Lambda}(\mathbf{x})=\arg\min\limits_{ \bm{\lambda}\in\Lambda}\|\bm{\lambda}-\mathbf{x}\|$. The Voronoi region of $\Lambda$ around $\mathbf{0}$, defined by $\mathcal{V}(\Lambda)=\{\mathbf{x}:Q_{\Lambda}(\mathbf{x})=\mathbf{0}\}$, specifies the nearest-neighbor decoding region. The Voronoi cell is an example of the fundamental region of the lattice, which is defined as a measurable set $\mathcal{R}(\Lambda)\subset\mathbb{R}^{n}$ if $\cup_{\bm{\lambda}\in\Lambda}(\mathcal{R}(\Lambda)+\bm{\lambda})=\mathbb{R}^{n}$ and if $(\mathcal{R}(\Lambda)+\bm{\lambda})\cap(\mathcal{R}(\Lambda)+\bm{\lambda}')$ has measure $\mathbf{0}$ for any $\bm{\lambda}\neq\bm{\lambda}'$ in $\Lambda$. The modulo lattice operation can be defined w.r.t. a fundamental region $\mathcal{R}$ as $\mathbf{x} \text{ mod}_{\mathcal{R}} \Lambda\triangleq \mathbf{x}-Q_{\mathcal{R}}(\mathbf{x})$, where $Q_{\mathcal{R}}({x})$ represents a lattice quantizer according to the region $\mathcal{R}$. The volume of a fundamental region is equal to that of the Voronoi region $\mathcal{V}(\Lambda)$, which is given by $V(\Lambda)=|\text{det}({\mathbf{B}})|$. The NVNR of an $n$-dimensional lattice $\Lambda$ is defined as $\gamma_{\Lambda}(\sigma)\triangleq V(\Lambda)^\frac{2}{n}/\sigma^2$. The NSM of $\Lambda$ is defined as $G(\Lambda) \triangleq \frac{1}{n V(\Lambda)}\cdot\frac{\int_{\mathcal{V}(\Lambda)}\|\mathbf{u}\|^2 d\mathbf{u}}{V^{2/n}(\Lambda)}$, where $\mathbf{u}$ is uniform in $\mathcal{V}(\Lambda)$.

\begin{deft}[AWGN-good lattices]\label{deft:AWGNgood}
A sequence of lattices $\Lambda_n$ of increasing dimension $n$ is modulation-good or AWGN-good, if for any fixed $P_{e}(\Lambda_n,\sigma^2)>0$ (defined in \eqref{eqn:LambdaPe}),
\begin{eqnarray}
\lim_{n\rightarrow\infty}\gamma_{\Lambda_n}(\sigma)=2\pi e. \notag\
\end{eqnarray}
\end{deft}

\begin{deft}[Quantization-good lattices]\label{deft:QZgood}
A sequence $\Lambda_n$ of lattices is called good for quantization or quantization-good under the mean square distortion measure if the NSM of $\Lambda_n$ satisfies
\begin{eqnarray}
\lim_{n \to \infty} G(\Lambda_n) = \frac{1}{2\pi e}.
\end{eqnarray}
\end{deft}

\begin{deft}\label{deft:Allgood}
A sequence $\Lambda_n$ of lattices is called simultaneously good if it is AWGN-good and quantization-good.
\end{deft}

A sublattice $\Lambda' \subset \Lambda$ induces a partition (denoted by $\Lambda/\Lambda'$) of $\Lambda$ into equivalence groups modulo $\Lambda'$. The order of the partition is equal to the number of cosets. We call it a binary partition if its order is 2. Let $\Lambda(\Lambda_0)/\Lambda_{1}/\cdots/\Lambda_{r-1}/\Lambda' (\Lambda_{r})$ for $r > 1$ be a 1-D lattice partition chain. To construct a multi-level lattice following the Construction D method \cite[p.232]{yellowbook}, for each partition $\Lambda_{\ell-1}/\Lambda_{\ell}$, a code $\mathscr{C}_{\ell}$ over $\Lambda_{\ell-1}/\Lambda_{\ell}$ selects a sequence of coset representatives $a_{\ell}$ in a set $A_{\ell}$ of representatives for the cosets of $\Lambda_{\ell}$. This construction requires a set of nested linear binary codes $\mathscr{C}_\ell$. When $\{\mathscr{C}_1,...,\mathscr{C}_r\}$ is a series of nested polar codes, we obtain a polar lattice \cite{yan2}. Let $\psi$ be the natural embedding of $\mathbb{F}_{2}^{N}$ into $\mathbb{Z}^{N}$, where $\mathbb{F}_{2}$ is the binary field. Consider $\mathbf{g}_{1},\mathbf{g}_{2},\cdots,\mathbf{g}_{N}$ be a basis of $\mathbb{F}_{2}^{N}$ such that $\mathbf{g}_{1},\cdots\mathbf{g}_{k_{\ell}}$ span $\mathscr{C}_{\ell}$, where $k_\ell$ is the dimension of $\mathscr{C}_{\ell}$.The binary lattice of Construction D consists of all vectors of the form $\sum_{\ell=1}^{r}2^{\ell-1}\sum_{j=1}^{k_{\ell}}u_{\ell}^{j}\psi(\mathbf{g}_{j})+2^{r}\mathbf{z}$, where $u_{\ell}^{j}\in\{0,1\}$ and $\mathbf{z}\in\mathbb{Z}^{N}$.

The standard Gaussian distribution of variance $\sigma^{2}$ centered at the origin is defined by
\begin{eqnarray}
f_{\sigma}(x)\triangleq\frac{1}{(\sqrt{2\pi}\sigma)^{n}}e^{-\frac{\| x\|^{2}}{2\sigma^{2}}}, \:\:x\in\mathbb{R}^{n}.
\end{eqnarray}

The differential entropy of $f_{\sigma}(x)$ is denoted by $\mathrm{h}(\sigma^2)$. For an AWGN channel with noise variance $\sigma^2$ per dimension, the probability of error $P_e(\Lambda, \sigma^2)$ of a minimum-distance decoder for $\Lambda$ is
\begin{eqnarray}\label{eqn:LambdaPe}
P_e(\Lambda, \sigma^2)=1-\int_{\mathcal{V}(\Lambda)} f_{\sigma}(x) dx.
\end{eqnarray}

The $\Lambda$-periodic function is defined as
\begin{eqnarray}\label{eq:LamP}
f_{\sigma,\Lambda}(x)\triangleq\sum\limits_{\lambda\in\Lambda}f_{\sigma,\lambda}(x)=\frac{1}{(\sqrt{2\pi}\sigma)^{n}}\sum\limits_{\lambda\in\Lambda}e^{-\frac{\| x-\lambda\|^{2}}{2\sigma^{2}}}.
\end{eqnarray}This distribution is actually the PDF of the $\Lambda$-aliased Gaussian noise, i.e., the Gaussian noise after the mod-$\Lambda$ operation \cite{forney6}. In this sense, the differential entropy of the $\Lambda$-aliased Gaussian noise is defined by
\begin{eqnarray}
\mathrm{h}(\Lambda, \sigma^2) \triangleq -\int_{\mathcal{V}(\Lambda)} f_{\sigma,\Lambda}(x) \log f_{\sigma,\Lambda}(x) dx.
\end{eqnarray}

\section{Simultaneously Good Polar Codes}\label{sec:2goodPCs}
\subsection{Two Good Codes with Same Rate}\label{sec:test}
To construct the simultaneously good polar codes, we start with two polar codes with the same length and rate that are good for channel coding and source coding, respectively. To keep the description simple, we restrict ourselves to binary symmetric channels (BSCs) and binary symmetric sources (BSSs) in this section. Let $W$ and $V$ be two BSCs with crossover probabilities $p$ and $p+\delta$, denoted by $BSC(p)$ and $BSC(p+\delta)$, respectively.
\begin{lem}\label{lem:2good}
For any small $\delta>0$ and $0<p<p+\delta<\frac{1}{2}$, there exist two good polar codes $\mathscr{P}_1$ and $\mathscr{P}_2$ with the same block length $N$ and rate $I(V)<\frac{K}{N} < I(W)$ such that $\mathscr{P}_1$ guarantees a decoding error probability less than $N2^{-N^{\beta}}$ as a channel code for $W$, and $\mathscr{P}_2$ achieves an average Hamming distortion less than $p+\delta+O(2^{-N^{\beta}})$ as a source code for a BSS.
\end{lem}
\begin{IEEEproof}
By the capacity-achieving property of polar codes, there exist a coding rate $R_c<I(W)$ and a block length $N_c$ satisfying the following condition.
\begin{eqnarray}\label{eqn:cset}
\left|\{i\in [N_c]: Z(W_{N_c}^{(i)}) < 2^{-N_c^\beta}\}\right| \geq R_cN_c.
\end{eqnarray}
Similarly, by the optimality of polar codes for lossy source coding, there exist a coding rate $R_s>I(V)$ and a block length $N_s$ satisfying the following condition.
\begin{eqnarray}\label{eqn:sset}
\left|\{i\in [N_s]: Z(V_{N_s}^{(i)}) < 1- 2^{-N_s^\beta}\right| \leq R_sN_s.
\end{eqnarray}
Let $N=\max\{N_c,N_s\}$. The above two conditions can be met with a common block length $N$. Since $I(V)<I(W)$ for any $\delta>0$, we can assume $R_s \leq R_c$ without loss of generality by making $N$ sufficiently large. Now choose the information set $\mathcal{I}_s=\{i\in [N]: Z(V_{N}^{(i)}) < 1- 2^{-N^\beta}\}$ for $\mathscr{P}_2$ and let $K=|\mathcal{I}_s|$. One can prove that $\mathscr{P}_2$ achieves expected distortion less than $p+\delta+O(2^{-N^{\beta}})$ following the steps of \cite[Thm. 3.4]{polarchannelandsource}. Next, let $\mathcal{I}_c$ be composed of the best $K$ indices (with the smallest Bhattacharyya parameters) in $\{i\in [N]: Z(W_{N}^{(i)}) < 2^{-N^\beta}\}$, and construct $\mathscr{P}_1$ according to $\mathcal{I}_c$. Using \cite[Prop. 2]{arikan2009channel}, $\mathscr{P}_1$ induces an error probability less than $N2^{-N^{\beta}}$ under SC decoding.
\end{IEEEproof}
\begin{rem}
The parameters $R_c$, $R_s$, $I(W)$ and $I(V)$ are demonstrated in Fig. \ref{fig:2good}. Note that Lemma \ref{lem:2good} holds for a broader class of test channels such as binary erasure channels and binary input AWGN channels, as long as the source $Y$ and the distortion function $\mathrm{d}(\cdot,\cdot)$ are adaptively matched.
\begin{figure}[ht]
    \centering
    \includegraphics[width=0.8\linewidth]{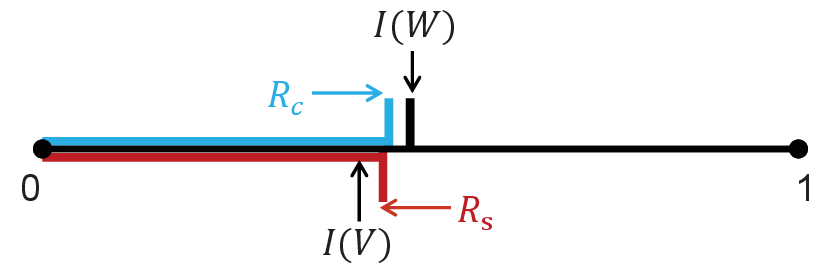}
    \caption{The constructions of $\mathscr{P}_1$ and $\mathscr{P}_2$ in Lemma \ref{lem:2good}.}
    \label{fig:2good}
\end{figure}
\end{rem}

\subsection{Chaining to a Single Long Code}\label{sec:chainPolar}
\begin{figure}[ht]
    \centering
    \includegraphics[width=0.95\linewidth]{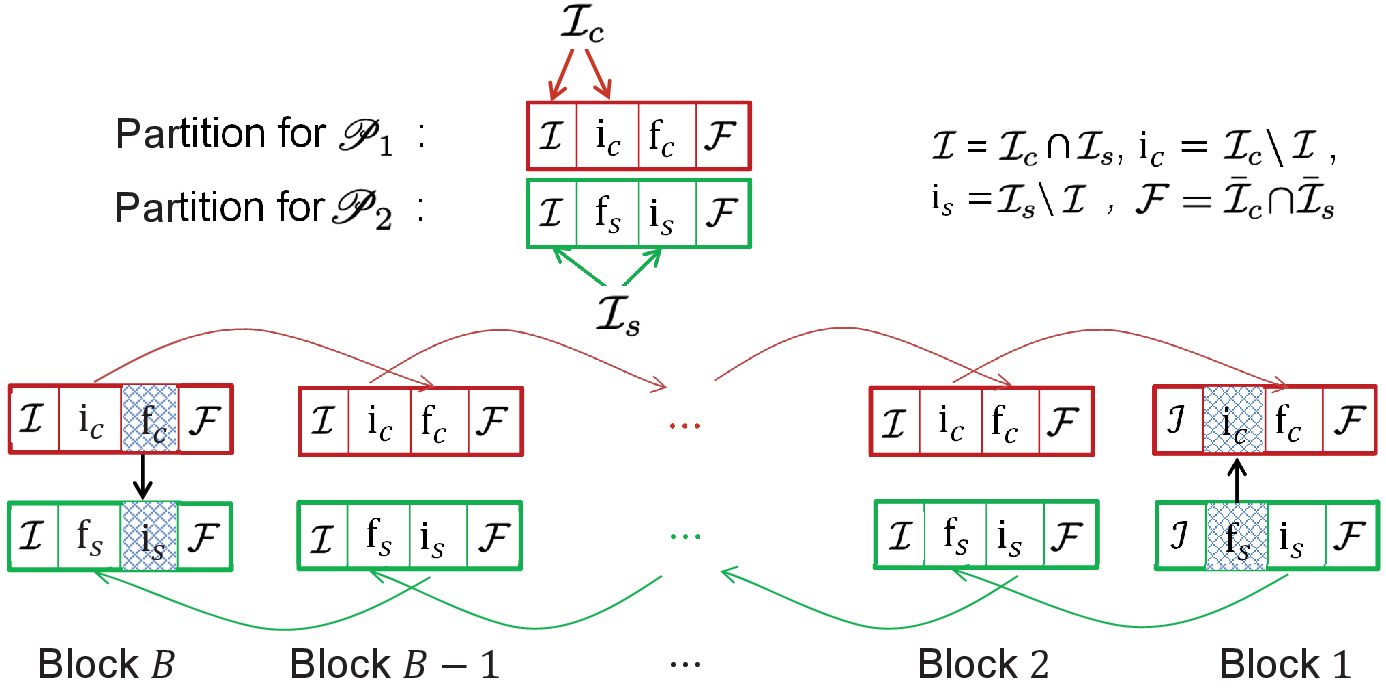}
    \caption{The structure of the chained polar code $\widehat{\mathscr{P}}$. The set partition marked in red is for channel coding, while the partition in green is for source coding. The encoding for channel coding is performed from the left to the right, with the block index decreases from $B$ to $1$. The set $\mathtt{i}_c^1$ in the 1-st block is fed with frozen bits instead of message bits; The encoding for source coding is performed reversely from Block 1 to $B$. For the last block, the set $\mathtt{i}_s^B$ is not determined by the source vector, but is fed with frozen bits. The arrows between the shadowed blocks highlight these special treatments.}
    \label{fig:ChainPolar}
\end{figure}

Now we chain $\mathscr{P}_1$ and $\mathscr{P}_2$ together to construct a simultaneously good code. Note that although the information sets $\mathcal{I}_c$ and $\mathcal{I}_s$ have the same size, their elements are not necessarily the same. Define $\mathcal{I}=\mathcal{I}_c\cap\mathcal{I}_s$, $\mathtt{i}_c =\mathtt{f}_s= \mathcal{I}_c\setminus\mathcal{I}$, $\mathtt{i}_s =\mathtt{f}_c = \mathcal{I}_s\setminus\mathcal{I}$, and $\mathcal{F}=\bar{\mathcal{I}}_c\cap\bar{\mathcal{I}}_s$. Clearly $\mathcal{I}\cup\mathtt{i}_c\cup\mathtt{i}_s\cup\mathcal{F} = [N]$. The set partitions are depicted in Fig. \ref{fig:ChainPolar}. It can be seen that when $\delta$ is small, the two sets $\mathcal{I}_c$ and $\mathcal{I}_s$ are close, and their intersection makes up the majority. The following lemma summarizes this result.
\begin{lem}\label{lem:smalli}
Let $\mathcal{I}_c$ and $\mathcal{I}_s$ be the two information sets of $\mathscr{P}_1$ and $\mathscr{P}_2$, respectively. For the set partition defined above, we have $|\mathcal{I}_c|=|\mathcal{I}_s|=K$, and $|\mathtt{i}_c| =|\mathtt{f}_s| = |\mathtt{i}_s| =|\mathtt{f}_c|$. Moreover,
\begin{eqnarray}
\lim_{N\to\infty}\frac{|\mathtt{i}_c|}{N}=0 \text{ and } \lim_{N\to\infty} \frac{|\mathcal{I}|}{N} = \frac{K}{N}.
\end{eqnarray}
\end{lem}
\begin{IEEEproof}
By the construction of $\mathscr{P}_1$ and $\mathscr{P}_2$, $|\mathcal{I}_c|=|\mathcal{I}_s|=K$. Therefore, $|\mathtt{i}_c| =|\mathtt{f}_s| = |\mathtt{i}_s| =|\mathtt{f}_c|$. Then $V$ is degraded by default w.r.t. $W$, which yields $Z(W_N^{(i)}) \leq Z(V_N^{(i)})$. For any $i \in \mathtt{i}_s$, $Z(V_N^{(i)}) < 1-2^{-N^\beta}$ and $Z(W_N^{(i)}) \geq 2^{-N^\beta}$, which means that $\mathtt{i}_s$ is a subset of the unpolarized set $\{i\in[N], 2^{-N^\beta}\leq Z(V_N^{(i)})<1-2^{-N^\beta}\}$. By channel polarization, $\lim_{N\to\infty}\frac{|\mathtt{i}_c|}{N}=0$.
\end{IEEEproof}

Now we define the chained polar code $\widehat{\mathscr{P}}$ based on $\mathscr{P}_1$ and $\mathscr{P}_2$.
\begin{deft}\label{deft:chainPolar}
Given the set partitions of $\mathscr{P}_1$ and $\mathscr{P}_2$, $\widehat{\mathscr{P}}$ is a code defined by the $B$ blocks of $\mathcal{I}, \mathtt{i}_c(\mathtt{f}_s), \mathtt{i}_s(\mathtt{f}_c)$ and $\mathcal{F}$, indexed by $1\leq j\leq B$ from the right end to the left. Its information set (for channel coding) is composed by $\mathcal{I}$ of the 1st block and $\mathcal{I}_s$ of the rest blocks, i.e.,
\begin{eqnarray}\label{eqn:cchain}
\widehat{\mathcal{I}}_c=\mathcal{I}^1\cup\mathcal{I}_c^2\cup\cdots\cup\mathcal{I}_c^B.
\end{eqnarray}
Or equivalently, its information set (for source coding) is composed by $\mathcal{I}$ of the $B$-th block and $\mathcal{I}_s$ of the rest blocks, i.e.,
\begin{eqnarray}\label{eqn:schain}
\widehat{\mathcal{I}}_s=\mathcal{I}^B\cup\mathcal{I}_s^{B-1}\cup\cdots\cup\mathcal{I}_s^1.
\end{eqnarray}
The $B$ blocks are chained by setting $U^{\mathtt{f}_c^j}=U^{\mathtt{i}_c^{j+1}}$, or $U^{\mathtt{f}_s^{j+1}}=U^{\mathtt{i}_s^{j}}$ for $1\leq j< B$. It can be seen that $\widehat{\mathcal{I}}_c$ and $\widehat{\mathcal{I}}_s$ generate the same codebook when $U^{\mathcal{F}}$'s are fixed for all blocks. The structure of $\widehat{\mathscr{P}}$ is shown in Fig. \ref{fig:ChainPolar}.
\end{deft}

\begin{ther}\label{thm:goodPCs}
The chained polar code $\widehat{\mathscr{P}}$ has block length $BN$ and rate $\frac{K}{N}-\frac{|\mathtt{i}_c|}{BN}$. For any $\delta>0$ and a sufficiently large $N$ such that $I(V)<\frac{K}{N}<I(W)$ holds,  $\widehat{\mathscr{P}}$ is simultaneously good in the sense that when $\widehat{\mathscr{P}}$ is employed for channel coding over $W$, the block error probability of $\widehat{\mathscr{P}}$ under SC decoding is upper-bounded by
\begin{eqnarray}\label{eqn:cgood}
P_e^{SC}(\widehat{\mathscr{P}}) \leq B\cdot N \cdot 2^{-N^\beta},
\end{eqnarray}and when $\widehat{\mathscr{P}}$ is utilized to compress a BSS with target average distortion $p+\delta$, the realized average distortion under SC decoding is upper-bounded by
\begin{eqnarray}
D^{SC}_{\rm{avr}}(\widehat{\mathscr{P}}) \leq p+\delta+4d_{\rm{max}}\cdot\frac{|\mathtt{i}_s|}{B},
\end{eqnarray}where $d_{\rm{max}}$ is the maximum value of $\mathrm{d}(\cdot,\cdot)$.
\end{ther}

\begin{IEEEproof}
For simplicity, we first assume that the frozen bits in $\mathtt{f}_s^1$, $\mathtt{f}_c^B$ and all $\mathcal{F}$'s are uniformly random and preshared with the receiver before transmission or compression. For channel coding, random message bits are assigned to $U^{\widehat{\mathcal{I}}_c}$. Then, $U^{\mathtt{f}_c^{j}}$ is copied from $U^{\mathtt{i}_c^{j+1}}$ for $1\leq j<B$. The message possesses $KB-|\mathtt{i}_c|$ bits since only $\mathtt{i}_c^1$ is wasted. The encoded codeword $X^{BN}$ is obtained by performing polar encoding for each block. After receiving $Y^{BN}$, the decoder recovers the message bits from the left end to the right. Using SC decoding, each block can be correctly decoded with an error probability less than $N\cdot2^{-N^\beta}$. Therefore, by the union bound, \eqref{eqn:cgood} is proved.

Let $\widehat{\mathcal{F}}_s$ be the complement set of $\widehat{\mathcal{I}}_s$, i.e., $\widehat{\mathcal{F}}_s=[BN]\setminus \widehat{\mathcal{I}}_s$. With some abuse of notation, let $Y^{BN}$ denote the source vector generated by a BSS. Consider compressing $Y^{BN}$ to $U^{\widehat{\mathcal{I}}_s}$ when $U^{\widehat{\mathcal{F}}_s}$ is randomly generated and independent of $Y^{BN}$. Similarly to \cite{KoradaSource}, we assume that $U^{\widehat{\mathcal{I}}_s}$ is generated from $Y^{BN}$ and $U^{\widehat{\mathcal{F}}_s}$ by random rounding instead of hard decision, following the SC decoding schedule. Let $Q^B_{\mathbf{U},\mathbf{Y}}$ denote the joint distribution between $Y^{(B-1)N+1:BN}$ and $U^{(B-1)N+1:BN}$ in the last block after compression. Let $P^{B}_{\mathbf{U},\mathbf{Y}}$ denote the joint distribution without compression, that is, all bits in $U^{(B-1)N+1:BN}$ are generated according to the SC decoding rule. The total variation distance $\mathbb{V}(P^B,Q^B)$ between $Q^B_{\mathbf{U},\mathbf{Y}}$ and $P^B_{\mathbf{U},\mathbf{Y}}$ can be upper-bounded as \eqref{eqn:chainTV}, where $\mathbb{D}(\cdot||\cdot)$ is the Kullback-Leibler divergence, and the equalities and the inequalities follow from
\begin{itemize}
\item[] $(a)$ $Q\left(u^i|u^{1:i-1},\mathbf{y}\right)=P\left(u^i|u^{1:i-1},\mathbf{y}\right)$ for $i \in \mathcal{I}^B$.
\item[] $(b)$ Pinsker's inequality.
\item[] $(c)$ Jensen's inequality.
\item[] $(d)$ $Q\left(u^i|u^{1:i-1}\right)=\frac{1}{2}$ for $i \in \mathcal{F}^B \cup \mathtt{i}_s^B$, and $U^{\mathtt{f}_s^B}$ is copied from $U^{\mathtt{i}_s^{B-1}}$ of the $(B-1)$-th block which is independent of $\mathbf{Y}$.
\item[] $(e)$ $Z(X|Y)^2<H(X|Y)$ \cite{polarsource}.
\item[] $(f)$ Definitions of $\mathcal{F}^B$ and $\mathtt{f}_s^B$.
\end{itemize}
For \eqref{eqn:chainTV}, we can see that the term $N\sqrt{4\ln2\cdot2^{-N^\beta}}$ is caused by $\mathcal{F}^B \cup \mathtt{f}_s^B$, and $|\mathtt{i}_s|\sqrt{2\ln2}$ is due to $\mathtt{i}_s^B$. For other blocks, we can check that only the first term exists. Notice that $V$ is the test channel, and $P^{1:B}_{\mathbf{U},\mathbf{Y}}$ induces a total distortion $BN(p+\delta)$. Therefore, the additional distortion induced by $Q^{1:B}_{\mathbf{U},\mathbf{Y}}$ is upper-bounded by
\begin{eqnarray}
\begin{aligned}
D^{SC}_{\rm{add}}&(\widehat{\mathscr{P}}) \\
&\leq BN^2d_{\rm{max}}\sqrt{4\ln2\cdot2^{-N^\beta}} +Nd_{\rm{max}}\cdot|\mathtt{i}_s|\sqrt{2\ln2} \notag\\
&\leq  4 N d_{\rm{max}} |\mathtt{i}_s|,
\end{aligned}
\end{eqnarray}
because $|\mathtt{i}_s|= o(N)$ by the finite scaling of polar codes \cite{HassaniScal14}. The proof is complete.
\end{IEEEproof}
\begin{rem}
By the symmetric settings of BMSC and BSS, the choice of frozen bits does not matter to the performance of channel coding and source coding \cite{arikan2009channel,KoradaSource}. Therefore, the bits in $\mathtt{f}_s^1$, $\mathtt{f}_c^B$ and all $\mathcal{F}$'s can be fixed as zeros for convenience. Note that the problematic set $\mathtt{i}_s^B$ is placed in the last block and only contaminates the last $N$ reconstruction symbols. The extra distortion caused is eventually absorbed by averaging over the block length $BN$. Notice that the neighboring blocks are linked by letting $U^{\mathtt{f}^{j+1}_s} = U^{\mathtt{i}^j_s}$ or $U^{\mathtt{f}^{j}_c} = U^{\mathtt{i}^{j+1}_c}$. To fully remove the statistical relationship between $Q^j_{\mathbf{U},\mathbf{Y}}$ and $Q^{j+1}_{\mathbf{U},\mathbf{Y}}$ for easier analysis, we may further assume that a mask sequence with uniformly random bits is used for each block. That is, $U^{\mathtt{f}^{j+1}_s} = U^{\mathtt{i}^j_s} \oplus R^{\mathtt{f}^{j+1}_s}$, where $R^{\mathtt{f}^{j+1}_s}$ denotes the random mask pre-shared between the two sides.  
\end{rem}

\begin{figure}[ht]
    \centering
    \includegraphics[width=0.8\linewidth]{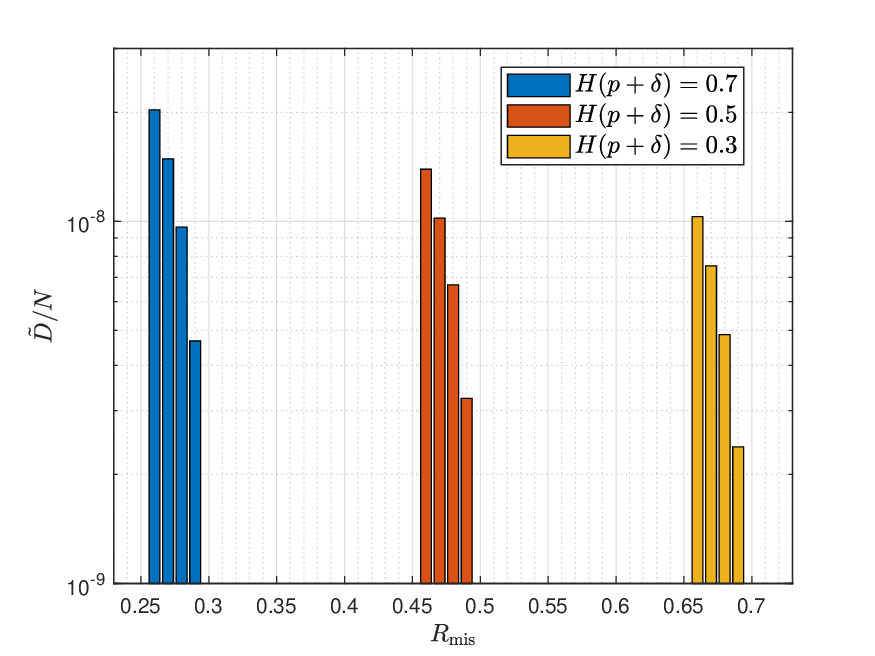}
    \vspace{-1em}
    \caption{The average extra distortion caused by $\mathtt{i}_s^B$ with different sizes.}
    \label{fig:n20_chain}
\end{figure}

\begin{rem}
The scaling of the extra distortion is much worse than the original polar code $\mathscr{P}_2$ due to the problematic set $\mathtt{i}_s^B$. By Lemma \ref{lem:smalli}, setting $B=O(N^t)$ for $t>1$ can still make $\widehat{\mathscr{P}}$ achieve the rate-distortion bound. In Fig. \ref{fig:n20_chain}, we show that in practice the influence of $\mathtt{i}_s^B$ is insignificant by simulation. To achieve the distortion of $p+\delta$,  the compression rate should be asymptotically equal to $R=1-H(p+\delta)$, where $H(\cdot)$ is the binary entropy function. If we use a mismatched rate of $R_{\rm{mis}}=1-H(p+\delta)- \tilde{R}$, an extra distortion $\tilde{D}$ is generated compared with the case of $R=1-H(p+\delta)$. If $\tilde{D}$ is normalized by $N^t$, it can approximately reflect the extra distortion of the chained polar code consisting of $N^t$ blocks. Strictly speaking, $\tilde{R}$ should be equal to $\frac{|\mathtt{i}_s|}{N}$. However, to roughly evaluate the effect of $|\mathtt{i}_s|$, we assume that $\tilde{R}$ takes some small values. In \mbox{Fig. \ref{fig:n20_chain}}, we show the relationship between the extra distortion $\tilde{D}$ normalized by $N$ and the mismatched rate $R_{\rm{mis}}$. The parameters are set as $N=2^{20}$, $H(p+\delta)=0.3, 0.5, 0.7$, and $\tilde{R} =0.01, 0.02, 0.03, 0.04$. It can be seen that the rate gap $\tilde{R}$ has a negligible effect on the average distortion.
\end{rem}

% \begin{rem}
% The scaling of the extra distortion is much worse than the original polar code $\mathscr{P}_2$ due to the problematic set $\mathtt{i}_s^B$. By Lemma \ref{lem:smalli}, setting $B=O(N^t)$ for $t>1$ can still make $\widehat{\mathscr{P}}$ achieve the rate-distortion bound. In Fig. \ref{fig:n20_chain}, we show that in practice the influence of $\mathtt{i}_s^B$ is insignificant by simulation. To achieve the distortion of $p+\delta$,  the compression rate should be asymptotically equal to $R=1-H(p+\delta)$, where $H(\cdot)$ is the binary entropy function. If we use a mismatched rate of $R=1-H(p+\delta)- \frac{|\mathtt{i}_s|}{N}$, an extra distortion $\tilde{D}$ is generated. If $\tilde{D}$ is normalized by $N^t$, it can approximately reflect the extra distortion of the chained polar code consisting of $N^t$ blocks. In \mbox{Fig. \ref{fig:n20_chain}}, we show the extra distortion $\tilde{D}$ normalized by $N$ for $H(p+\delta)=0.3, 0.5, 0.7$ and $\frac{|\mathtt{i}_s|}{N} =0.01, 0.02, 0.03, 0.04$. It can be seen that the rate gap $\frac{|\mathtt{i}_s|}{N}$ has a negligible effect on the average distortion.
% \end{rem}

\section{Simultaneously Good Polar Lattices}\label{sec:GoodPL}
The simultaneously good polar lattices can now be constructed from simultaneously good polar codes introduced in the previous section. With some abuse of notation, let $W$ and $V$ denote two AWGN channels, with noise variances $\sigma^2$ and $\sigma^2+\delta$, respectively. Given a 1-D binary partition chain $\Lambda(\Lambda_0)/\Lambda_{1}/\cdots/\Lambda_{r-1}/\Lambda' (\Lambda_{r})$, both $W$ and $V$ are decomposed into a series of binary partition channels, indicated by $W(\Lambda_{\ell-1}/\Lambda_{\ell},\sigma^2)$ and $V(\Lambda_{\ell-1}/\Lambda_{\ell},\sigma^2+\delta)$ at the level $\ell$, respectively. Clearly, $V(\Lambda_{\ell-1}/\Lambda_{\ell},\sigma^2+\delta)$ is degraded w.r.t. $W(\Lambda_{\ell-1}/\Lambda_{\ell},\sigma^2)$, and a chained polar code $\widehat{\mathscr{P}}_\ell$ can be constructed by following the same method given in Sec. \ref{sec:2goodPCs}. The next lemma guarantees that such chained polar codes are sequentially nested.
\begin{lem}
$\widehat{\mathscr{P}}_\ell \subseteq \widehat{\mathscr{P}}_{\ell+1}$ for $1\leq \ell < r$. Therefore, the codes $\{\widehat{\mathscr{P}}_1,...,\widehat{\mathscr{P}}_r\}$ form a valid polar lattice with $r$ levels by Construction D.
\end{lem}
\begin{IEEEproof}
It is sufficient to prove that the information set $\widehat{I}_\ell$ at level $\ell$ is a subset of $\widehat{I}_{\ell+1}$ at level $\ell+1$. Note that both $\widehat{I}_\ell$ and $\widehat{I}_{\ell+1}$ can be expressed as \eqref{eqn:cchain} for channel coding or \eqref{eqn:schain} for source coding. According to the result in \cite{polarlatticeJ}, $W(\Lambda_{\ell-1}/\Lambda_{\ell},\sigma^2)$ ($V(\Lambda_{\ell-1}/\Lambda_{\ell},\sigma^2+\delta)$) is degraded w.r.t. $W(\Lambda_{\ell}/\Lambda_{\ell+1},\sigma^2)$ ($V(\Lambda_{\ell}/\Lambda_{\ell+1},\sigma^2+\delta)$), which means that the information sets $\mathcal{I}_c$ and $\mathcal{I}_s$ of the component polar codes at level $\ell$ are the subsets of those at level $\ell+1$, respectively. Moreover, this relationship also holds for the intersection set $\mathcal{I}=\mathcal{I}_c \cup \mathcal{I}_s$. Therefore, by Definition \ref{deft:chainPolar}, $\widehat{I}_\ell \subseteq  \widehat{I}_{\ell+1}$, and $\widehat{\mathscr{P}}_\ell \subseteq \widehat{\mathscr{P}}_{\ell+1}$.
\end{IEEEproof}

Let $\widehat{\Lambda}$ denote the polar lattice constructed from the simultaneously good polar codes. We can choose a common \mbox{1-D} binary partition $\Lambda_0/\Lambda_{1}/\cdots/\Lambda_{r-1}/\Lambda_{r}$ for both $W$ and $V$ such that the \text{mod-}$\Lambda_0$ channel capacity $C(\Lambda_0, \sigma^2)\to 0$ and the bottom lattice $\Lambda_r$ has negligible effect on the Gaussian noise, i.e., $\mathrm{h}(\Lambda_r,\sigma^2+\delta) \approx \mathrm{h}(\sigma^2+\delta)$. According to \cite[Prop. 1]{QZgoodArxiv}, setting $r=O(\log N)$ guarantees such fine top lattice $\Lambda_0$ and coarse bottom lattice $\Lambda_r$, which has volume $\sqrt{N}$. By the continuity of the differential entropy function, $\delta$ induces a constant gap in capacity $\Delta(\delta)\triangleq C(\Lambda_r,\sigma^2)-C(\Lambda_r,\sigma^2+\delta)$, which means that the gaps between the capacities of the partition channels $W(\Lambda_{\ell}/\Lambda_{\ell+1},\sigma^2)$ and $V(\Lambda_{\ell}/\Lambda_{\ell+1},\sigma^2+\delta)$ can be set as $\frac{\Delta(\delta)}{r} = \Omega(\frac{1}{\log N})$. In this case, Lemma \ref{lem:2good} should be modified for a vanishing capacity gap $\Omega(\frac{1}{\log N})$ instead of an arbitrary constant. To do so, we turn the finite length scaling of binary polar codes \cite{HassaniScal14}. For a given block error probability $P_e >0$, the gap between the coding rate and the capacity scales like $N^{-\frac{1}{\mu}}$, where $\mu>2$ is called \emph{scaling exponent}. For sufficiently large $N$, $N^{-\frac{1}{\mu}}$ vanishes faster than $\frac{1}{\log N}$, meaning that the two basic polar codes in Lemma \ref{lem:2good} can still be constructed, with a penalty on the error exponent $\beta$. Roughly speaking, the joint scaling law says that $Z(W_N^{(i)})$ in \eqref{eqn:cset} behaves as $e^{-\alpha N^{\frac{\kappa}{\mu}}(I(W)-R_c)^{\kappa}}$ for some positive constants $\alpha$ and $\kappa$. When $I(W)-R_c$ is fixed, $\frac{\kappa}{\mu}$ can be close to any $\beta<\frac{1}{2}$ as proved in \cite{arikan2009rate}. When $I(W)-R_c$ is required to vanish as $\Omega(\frac{1}{\log N})$, there exists a $0<\beta'<\beta<\frac{1}{2}$ such that \eqref{eqn:cset} holds with $\beta$ replaced by $\beta'$. Similarly, by the symmetry between the polarization around 0 and 1, such $\beta'$ can be found for \eqref{eqn:sset} as well. A detailed description of the joint scaling law is given in \cite[Sec. IV]{MondelliScal16}. Therefore, for each partition level $\ell$, a simultaneously good polar code $\widehat{\mathscr{P}_\ell}$ can be constructed with error exponent $\beta'$, as in Theorem \ref{thm:goodPCs}. The $BN$-dimensional lattice $\widehat{\Lambda}$ can be constructed. 

\begin{ther}
The sequence of $\widehat{\Lambda}$ with dimension $BN$ is simultaneously good.
\end{ther}
\begin{IEEEproof}
We only provide a sketch of the proof due to limited space. The AWGN-goodness can be proved similarly to that in \cite{polarlatticeJ}. It is not difficult to see that the normalized volume of $\widehat{\Lambda}$ is close to $2\pi e\sigma^2$ as the coding rate $R_{\ell}$ of $\widehat{\mathscr{P}_\ell}$ approaches $C(\Lambda_{\ell-1}/\Lambda_{\ell},\sigma^2)$. That is,
\begin{eqnarray}\label{eqn:vnr}
\begin{aligned}
\log\left(\frac{\gamma_{\widehat{\Lambda}}(\sigma)}{2\pi e}\right) &= \log \frac{V(\widehat{\Lambda})^{\frac{2}{BN}}}{2\pi e \sigma^2} \\
&= \log \frac{2^{-2R_{\mathcal{C}}}V(\Lambda')^2}{2\pi e \sigma^2} \\
& = -2R_{\mathcal{C}} + 2 \log V(\Lambda') - \log(2\pi e \sigma^2),
\end{aligned}
\end{eqnarray}Define
\begin{eqnarray}
\begin{cases} \epsilon_{a}=C(\Lambda,\sigma^2) \\
\epsilon_{b}=\mathrm{h}(\sigma^2)-\mathrm{h}(\Lambda',\sigma^2) \\\notag
\epsilon_{c}=C(\Lambda/\Lambda', \sigma^2)-R_{\mathcal{C}}=\sum_{\ell=1}^{r}{C(\Lambda_{\ell-1}/\Lambda_{\ell}, \sigma^2)}-R_{\ell}.
\end{cases}
\label{eqn:epsilons}
\end{eqnarray}
We have $\log\left(\frac{\gamma_{\widehat{\Lambda}}(\sigma)}{2\pi e}\right)=2(\epsilon_a - \epsilon_b + \epsilon_c)$. By the construction of the partition chain, $\epsilon_a \to 0$, $\epsilon_b>0$ and $\epsilon_c \leq \Delta(\delta)$. Therefore, $\log\left(\frac{\gamma_{\widehat{\Lambda}}(\sigma)}{2\pi e}\right) \leq 4\Delta(\delta)$ for any $\delta>0$ and sufficiently large $N$. By the union bound, the decoding error probability can be bounded as
\begin{eqnarray}
\begin{aligned}
P_e^{SC}(\widehat{\Lambda}) &\leq rBN \cdot 2^{-N^{\beta'}} + P_e(\Lambda'^{BN},\sigma^2) \\
&\leq   rBN \cdot 2^{-N^{\beta'}} + BN \cdot P_e(\Lambda', \sigma^2) \\
&\leq  rBN \cdot 2^{-N^{\beta'}} + 2BN \cdot \exp\left(-\frac{N}{8\sigma^2}\right).
\end{aligned}
\end{eqnarray}

The proof of quantization-goodness is more complex, and it follows the steps of our recent work \cite{QZgoodITW2024}. Similarly to \cite[Thm. 2]{QZgoodArxiv}, the average distortion achieved by $\widehat{\Lambda}$ is upper-bounded by
\begin{eqnarray}
\frac{1}{BN} \mathsf{E}\left[\left\|\alpha Y^{[BN]}-X^{[BN]}\right\|^2\right]\leq \sigma^2 +\delta + \frac{2r|\mathtt{i}_s|_{\rm{max}}}{B},
\end{eqnarray}where $\alpha$ is the minimum mean square error (MMSE) re-scaling factor, $\sigma^2 +\delta$ is the target distortion and $|\mathtt{i}_s|_{\rm{max}}$ is the maximum size of $\mathtt{i}_s$ among all partition levels. We see the main difference is that the sub-exponentially vanishing term $N\cdot 2^{-N^{\beta'}}$ is replaced with $\frac{r|\mathtt{i}_s|_{\rm{max}}}{B}$, which decays polynominally by letting $B=O(N^t)$ for $t>1$. The next step is to show that the variance of the average distortion over all random mapping is vanishing, which means that $\frac{1}{BN} \mathsf{E}\left[\left\|\alpha Y^{[BN]}-X^{[BN]}\right\|^2\right]$ is bounded by a radius slightly larger than $\sigma^2+\delta$ with high probability. In a similar fashion of \cite[Lem. 5]{QZgoodArxiv}, we can prove that the second moment $\sigma^2(\widehat{\Lambda})$ satisfies
\begin{eqnarray}
\sigma^2(\widehat{\Lambda}) \leq \frac{N}{N+2}(\sigma^2+\delta+o(N^{-\tau})),
\end{eqnarray}where $\tau=\frac{2}{3}(t-2)$. A safe choice is to have $B=N^3$ ($t=3$) and $\tau=\frac{2}{3}$.

Finally, by bounding the volume $V(\widehat{\Lambda})$ from the other side in \eqref{eqn:vnr}, we see that $V(\widehat{\Lambda})^{\frac{2}{BN}}$ is indeed close to $2\pi e (\sigma^2+\delta)$, which eventually gives us $\lim_{N\to \infty} G(\widehat{\Lambda})=\frac{1}{2\pi e}$. More details are available in \cite{QZgoodArxiv} and \cite{QZgoodITW2024}.
\end{IEEEproof}

\section{Conclusion}
In this work, we found an explicit construction of simultaneously good polar codes and polar lattices. For future work, we may compare the scaling parameters of our polar lattices with the random lattice ensembles in \cite{zamir2}. The convergence rates of the NVNR and the NSM of $\widehat{\Lambda}$ are also worth researching. 

\begin{figure*}[th]
\begin{eqnarray}\label{eqn:chainTV}
\begin{aligned}
&2\mathbb{V}\left(P^B_{\mathbf{U},\mathbf{Y}},Q^B_{\mathbf{U},\mathbf{Y}}\right) \\
&= \sum_{\mathbf{u},\mathbf{y}} \left|Q(\mathbf{u},\mathbf{y})-P(\mathbf{u},\mathbf{y})\right|\\
&=\sum_{\mathbf{u},\mathbf{y}}\Bigg|\sum_{i=1}^{N}\left(Q(u^i|u^{1:i-1},\mathbf{y})-P(u^i|u^{1:i-1},\mathbf{y})\right)\cdot\left(\prod_{j=1}^{i-1}P(u^j|u^{1:j-1},\mathbf{y})\right)\left(\prod_{j=i+1}^{N}Q(u^j|u^{1:j-1},\mathbf{y})\right)P(\mathbf{y})\Bigg|\\
&\stackrel{(a)}\leq\sum_{i\in\overline{\mathcal{I}^B}}\sum_{\mathbf{u},\mathbf{y}}\left|Q(u^i|u_1^{1:i-1},\mathbf{y})-P(u^i|u^{1:i-1},\mathbf{y})\right|\left(\prod_{j=1}^{i-1}P(u^j|u^{1:j-1},\mathbf{y})\right)\cdot\left(\prod_{j=i+1}^{N}Q(u^j|u^{1:j-1},\mathbf{y})\right)P(\mathbf{y})\\
&=\sum_{i\in\overline{\mathcal{I}^B}}\sum_{u^{1:i},\mathbf{y}}\left|Q(u^i|u^{1:i-1},\mathbf{y})-P(u^i|u^{1:i-1},\mathbf{y})\right|\left(\prod_{j=1}^{i-1}P(u^j|u^{1:j-1},\mathbf{y})\right)P(\mathbf{y})\\
&=\sum_{i\in\overline{\mathcal{I}^B}} \sum_{u^{1:i-1},\mathbf{y}} 2P\left(u^{1:i-1},\mathbf{y}\right)\mathbb{V}\left(Q_{U^i|U^{1:i-1}=u^{1:i-1},\mathbf{Y}=\mathbf{y}},P_{U^i|U^{1:i-1}=u^{1:i-1},\mathbf{Y}=\mathbf{y}}\right)\\
&\stackrel{(b)}\leq \sum_{i\in\overline{\mathcal{I}^B}} \sum_{u^{1:i-1},\mathbf{y}} P\left(u^{1:i-1},\mathbf{y}\right) \sqrt{2\ln2 \mathbb{D}\left(P_{U^i|U^{1:i-1}=u^{1:i-1},\mathbf{Y}=\mathbf{y}}||Q_{U^i|U^{1:i-1}=u^{1:i-1},\mathbf{Y}=\mathbf{y}}\right)}\\
&\stackrel{(c)} \leq \sum_{i\in\overline{\mathcal{I}^B}} \sqrt{2\ln2 \sum_{u^{1:i-1},\mathbf{y}} P\left(u^{1:i-1},\mathbf{y}\right) \mathbb{D}\left(P_{U^i|U^{1:i-1}=u^{1:i-1},\mathbf{Y}=\mathbf{y}}||Q_{U^i|U^{1:i-1}=u^{1:i-1},\mathbf{Y}=\mathbf{y}}\right)}\\
&= \sum_{i\in\overline{\mathcal{I}^B}} \sqrt{2\ln2 \mathbb{D}\left(P_{U^i}||Q_{U^i}|U^{1:i-1},\mathbf{Y}\right)}\\
&\stackrel{(d)}= \sum_{i\in\overline{\mathcal{I}^B}} \sqrt{2\ln2\left(1-H(U^i|U^{1:i-1},\mathbf{Y})\right)}\\
&= \sum_{i\in\mathcal{F}^B\cup\mathtt{f}_s^B} \sqrt{2\ln2\left(1-H(U^i|U^{1:i-1},\mathbf{Y})\right)}+\sum_{i\in\mathtt{i}_s^B}\sqrt{2\ln2\left(1-H(U^i|U^{1:i-1},\mathbf{Y})\right)}\\
&\stackrel{(e)}\leq \sum_{i\in\mathcal{F}^B\cup\mathtt{f}_s^B} \sqrt{2\ln2\left(1-Z(U^i|U^{1:i-1},\mathbf{Y})^2\right)}+|\mathtt{i}_s|\sqrt{2\ln2}\\
&\stackrel{(f)}\leq N\sqrt{4\ln2\cdot2^{-N^\beta}}+|\mathtt{i}_s|\sqrt{2\ln2}\\
\end{aligned}
\end{eqnarray}
\hrulefill
\end{figure*}





%\bibliographystyle{IEEEtran}
\bibliographystyle{IEEEtran}
\bibliography{Myreff}

\end{document}